\documentclass{ws-ijmpe}

\begin{document}

\markboth{Liang \& Signorini}{Fusion induced by radioactive ion beams}

\catchline{}{}{}{}{}
%
%%%%%%%%%%%%%%%%%%%%%%%%%%%%%%%%%%%%%%%%%%%%%%%%%%%%%%%%%%%%%%%%%%%%

\title{FUSION INDUCED BY RADIOACTIVE ION BEAMS}

\author{J. F. Liang}

\address{Physics Division, Oak Ridge National Laboratory \\
Oak Ridge, Tennessee 37830, U.S.A.\\
liang@mail.phy.ornl.gov}

\author{C. Signorini}

\address{Physics Department of the University and INFN,\\
via Marzolo 8, 35120 Padova, Italy \\
cosimo.signorini@pd.infn.it}

\maketitle

\begin{history}
\received{(received date)}
\revised{(revised date)}
%\accepted{(Day Month Year)}
%\comby{(xxxxxxxxxx)}
\end{history}

\begin{abstract}
The use of radioactive beams opens a new frontier for fusion studies. The
coupling to the continuum can be explored with very loosely bound nuclei.
Experiments were performed with beams of nuclei at or near the proton and
neutron drip-lines to measure fusion and associated reactions in the
vicinity of the Coulomb barrier. In addition, the fusion yield is predicted
to be enhanced in reactions involving very neutron-rich unstable nuclei.
Experimental measurements were carried out to investigate if it is feasible
to use such beams to produce new heavy elements. The current status of these
experimental activities is given in this review.
\end{abstract}

%\section{Major Headings}
%Major headings should be typeset in boldface with the first
%letter of important words capitalized.
\section{Introduction}
The interaction of two nuclei consists of a repulsive Coulomb potential and
an attractive nuclear potential. The overlap of the two potentials produces
a Coulomb barrier. Fusion occurs when the interacting nuclei have sufficient
kinetic energy to overcome the repulsive barrier and are subsequently trapped
inside
the potential pocket to form a compound nucleus. The kinetic energy determines
whether the process takes place by going over or quantum tunneling through
the barrier. The nucleons
in the reactants are rearranged in the compound nucleus.

Nuclear fusion is responsible for energy generation in stars. It is also
a process
for synthesizing new elements in laboratories. The study of fusion has been
carried out for several decades. In the 1980s, the discovery of unexpectedly
large fusion cross sections at sub-barrier energies in some heavy-ion systems
generated tremendous interest in fusion studies.\cite{be85,st86,be88}
The sub-barrier fusion
enhancement can be understood in the coupled-channel formalism.\cite{ba98,da98}
The coupling
of the entrance channel to the intrinsic degrees of freedom of the projectile
and target splits the uncoupled single barrier into a distribution of barriers.
The enhanced fusion cross sections at sub-barrier energies arise from going
over the low energy barriers.

The advent of radioactive ion beams (RIBs) has generated new excitement
in this field.
The r.m.s. radius of nuclei far from the $\beta$-stability valley is, in many
cases, significantly larger. Fusion is expected to be enhanced due to lower
barriers.
However, these nuclei are often loosely bound, {\it i.e.}, the valence
nucleon(s) have very small binding energy. Breakup of the loosely bound nuclei
in the Coulomb and/or nuclear field of the target can
take place and thus remove the available flux for fusion. As a result, fusion
is predicted to be suppressed.\cite{hu92,ta93} On the other hand, if breakup
is significantly
large fusion can be enhanced through the strong coupling.\cite{da94}
How fusion is influenced by loosely bound nuclei is still an open question, as
will be seen later in this review.

Several stable beam experiments have showed that neutron transfer with positive
Q-values can
enhance sub-barrier fusion.\cite{mo94,st95,st97,ti97,so98}
With neutron-rich radioactive beams, the number of
transfer channels with positive Q-values can be very large. One would expect
an enhancement of sub-barrier fusion by the coupling to these transfer
reactions.\cite{de00,za03} The compound nucleus formed with neutron-rich
radioactive nuclei
induced reactions should have higher stability against fission. Using
short-lived neutron-rich radioactive beams may be a viable way for producing
new heavy elements.\cite{hu91} Results of fusion measurements performed
with some low
intensity neutron-rich radioactive beams will be discussed in this paper.
 
The emphasis of this review is on experimental work.
General discussions on recent theoretical developments in fusion can be found
in review articles by Balantekin and Takigawa,\cite{ba98}
Dasgupta {\it et al.},\cite{da98} and Signorini\cite{si01}. However,
specific theoretical treatments on some
of the measurements discussed will be presented where the experiment is
mentioned. The challenges in conducting fusion measurements with radioactive
beams will be shown in Sect. {\bf 2} followed by a brief description of
techniques for producing
radioactive beams. Measurements using light-mass and medium-mass radioactive
beams will be discussed in Sect. {\bf 4} and Sect. {\bf 5}, respectively.
Concluding remarks will be given in Sect. {\bf 6}.

\section{Experimental Considerations}
Fusion is commonly studied by measuring the cross section as a function of
reaction energy. The cross section can be determined by detecting the
evaporation residues or fission fragments directly, or by detecting the
$\gamma$ rays or light particles emitted from the evaporation residues
deposited in a catcher foil. The techniques used in radioactive beam
measurements are similar to those used in stable beam
experiments.\cite{be85,st86}

It is very important to have high quality beams for fusion measurements. Good
energy resolution, small energy width, is crucial for measuring excitation
functions, particularly at energies below the barrier where the cross
sections
change exponentially. To determine whether sub-barrier fusion is enhanced the
reaction energies have to be known better than $\sim$1\%. Low emittance beams
which
result in a beam spot of a few mm on target are desirable.
This reduces the angular broadening and energy spread of the reaction
products.
Beam purity is often an issue in RIBs, and it is associated with
the production processes. The unwanted contaminants can be removed or
reduced by high resolution mass separators or chemical methods. But sometimes
they cannot be completely eliminated because the beam of interest is
orders of magnitude less intense than the contaminants. Therefore,
measurements with the main contaminant beam species have to be performed and
subtracted from measurements with the mixed beam.

The intensity of RIBs is, presently, several orders of
magnitude lower than that of stable beams. In stable beam experiments, the beam
intensity can be measured by a Faraday cup. One
particle nA is $6.24\times 10^{9}$ particles per second
(pps). In RIB experiments, an intensity of $10^{6}$ to $10^{7}$ pps would be
considered high today.
Fusion measurements can be performed with a beam of 10$^{4}$ pps.
In order to compensate for the low beam intensity, high efficiency
detectors with large solid angle coverage or multiple targets for obtaining
measurements of several energies in one run have been employed. The
availability of large area silicon strip detectors (SSD) in a variety shapes
and compact multichannel electronics for handling such detectors have made
many measurements feasible. On the other hand, because of the low beam
intensity, the use of event-by-event beam tracking
is very useful for cleaning up events originating from contaminant induced
reactions.
Incorporating time-of-flight measurements into experiments has many
advantages and is very common. It can be used to measure beam energies. When
beam contaminants are present, valid events can be selected using timing gates
corresponding to the correct beam particles. The decay of beam
particles sometimes contributes to the background in detectors. With the use
of time-of-flight, this background can be suppressed effectively.

\section{Types of Radioactive Ion Beams}
%Sub-headings should be typeset in boldface italic and capitalize
%the first letter of the first word only. Section number to be in
%boldface roman.
The production of RIBs has been discussed extensively in many
reports\cite{rnb6} and is beyond the scope of this review. We will give a brief
description of the methods used to produce beams for the experiments
discussed in this paper.
\subsection{Isotope separator on-line}
\begin{figure}[th]
\centerline{\psfig{file=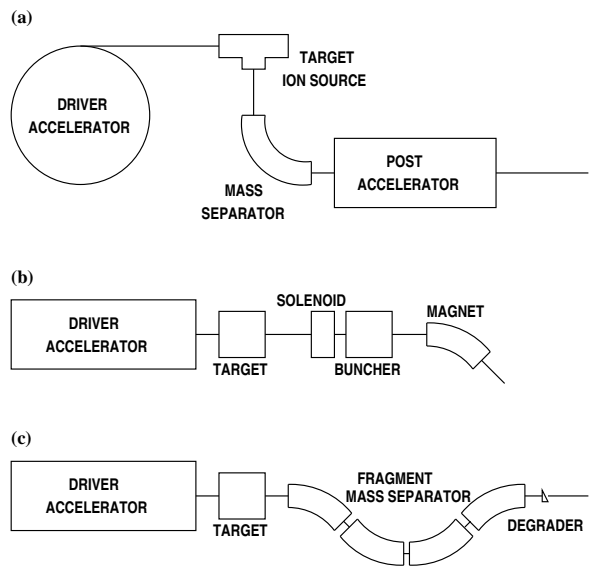,width=8cm}}
%\vspace*{8pt}
\caption{A schematic illustration of RIB production: (a) ISOL, (b) in-flight
with direct reactions, and (c) in-flight with fragmentation reactions.
\label{fg:ribprod}}
\end{figure}

%Typeset sub-subheadings in medium face italic and capitalize the
%first letter of the first word only. Section numbers to be in
%roman.
The isotope separator on-line (ISOL) method uses a driver accelerator to
accelerate light-charged particles, such as protons, deuterons, and $\alpha$
particles, into a thick target.\cite{st00,vi01,ry02} The
radioactive atoms diffuse out of the target and get ionized in an
ion source. The secondary ions are then selected by mass separators,
accelerated by
a post accelerator, and sent to an experimental area, as shown in
Fig.~\ref{fg:ribprod}(a). The beam quality is very
good because the post accelerator is usually an electrostatic tandem
accelerator or a linac. Beams which can be produced with this method are
limited to those with lifetimes of the order of seconds or longer.
Species with shorter lifetimes may not be able
to get out of the target fast enough for post acceleration.
How fast a radioactive atom gets out of the target
depends strongly on the chemical and structural properties of the target.
\subsection{In-flight}
In this method, radioactive beams can be produced with the use of one
accelerator. The accelerated heavy-ion beam is incident on a thin target in
inverse kinematics. The reaction products emitted in a narrow cone in the
forward direction are focused
and transported through mass separators. At energies near and below 10
MeV/nucleon, direct reactions are often used. The reaction products can be
selected and focused by solenoids such as the TWINSOL facility at the
University of Notre Dame,\cite{be97,le99}
or by a combination of solenoid, bunching resonator, and bending magnet
such as at ATLAS (Argonne National Laboratory).\cite{ha00}
Figure~\ref{fg:ribprod}(b) depicts the beam selection by the latter method.
At energies of a few tens
to several hundreds MeV/nucleon, the production reaction
is essentially fragmentation. Very sophisticated fragment mass separators such
as the
A1900 at NSCL (Michigan State University)\cite{mo03} and
RIPS at RIKEN\cite{ku92}, is used for
filtering the beam, as shown in Fig.~\ref{fg:ribprod}(c). The
advantage of the in-flight production method is that very short-lived
species can be produced because diffusion and target chemistry are not involved
in the processes. However, reaction mechanisms and target thickness
introduces significant energy and angular spreads resulting in poor energy 
resolution and large emittance of the secondary beam.

\section{Light-ion Reactions}
This section will discuss the fusion reactions induced by light-mass
radioactive ion beams namely, $^{6,8}$He, $^{11}$Be, $^{11}$C, and $^{17}$F
up to now,
which are the only projectiles utilized for this type of study. The $^{6,8}$He,
$^{11}$Be, and $^{17}$F projectiles are particularly interesting for the
following reasons:
\begin{itemize}
\item[a)] $^{6}$He is weakly bound with a two-neutron separation energy
S$_{2n}$ = 0.972 MeV and has a neutron
skin like structure with a large neutron spatial distribution
around the $^{4}$He core.\cite{ta92,al97} In the present literature $^{6}$He
is considered as a neutron halo nucleus. The isotope $^{8}$He is
more tightly bound with  S$_{2n}$ = 2.468 MeV but still has a structure similar
to $^{6}$He.\cite{ta92,al97}
\item[b)] $^{11}$Be is weakly bound, with S$_{n}$ = 0.504 MeV, and has a well
established neutron halo\cite{ta88,fu91}
which produces a r.m.s. radius $\sim$10\%
larger than what is expected from the $r_{0}A^{1/3}$ systematics with
$r_{0} \sim$ 1.18 fm.
\item[c)] $^{17}$F is a proton drip-line nucleus with a proton separation
energy S$_{p}$ = 0.600 MeV. Its
first excited state is bound by 0.105 MeV with an extended r.m.s. radius
$r_{rms}$ = 5.3 fm which is considered as a halo state.\cite{mo97}
\end{itemize}

In addition to the fusion process, the main topics of this review, discussion
also includes
the breakup (BU) process which is expected to be strong because of the small
binding energy of these short-lived nuclei. Such a process could
hinder or enhance the fusion process around the barrier as
extensively debated by many theoreticians, and/or increase the
total reaction cross section.

In the following subsections the experimental results obtained with the light
RIBs
mentioned above will be critically reviewed as  well as the theories developed
to explain the measurements.
Concerning fusion with loosely bound projectiles, which break up easily, we
have to distinguish between the following processes:
a) {\em complete fusion} (CF),
when all projectile nucleons are trapped inside the target, b)
{\em incomplete fusion} (ICF), when only part of the nucleons are
trapped, and c) {\em total fusion} (TF), the combination of
complete and incomplete fusion cross. In the absence of incomplete fusion,
total fusion and complete fusion are identical.

We should bear in mind that
theories usually calculate total fusion cross sections, while various
fusion cross sections measured by particular experiments.
In order to compare different systems, the fusion cross sections
should be corrected for geometric factors originating from the
different nuclear radii involved, and the energies of the colliding systems
(usually the center of mass energies) should be divided by the
Coulomb barrier energies, V$_{B}$. In the following, V$_{B}$ will be evaluated
with $r_{0} = 1.56$ fm. This value was deduced
from the barrier distribution for the system $^{9}$Be+$^{209}$Bi
($^{208}$Pb) in Ref.~\cite{si99,da99} with the standard formula
V$_{B}$~=~Z$_{p}$Z$_{t}$e$^{2}$/$r_{0}$(A$_{p}^{1/3}$+A$_{t}^{1/3}$)
for V$_{B}$ = 38.50 (37.75) MeV.

In connection with loosely bound nucleus induced fusion, there is a relevant
process where the valence nucleon of the projectile, such as the proton in
$^{17}$F, is captured by the target. Such a process can be a conventional
transfer/stripping and will be called ``transfer'', or a strong
capture/transfer/stripping breakup and will be called
``transfer/stripping-breakup''.
\subsection{$^{6,8}$He beams}
The $^{6}$He beam is becoming fairly popular. It has been produced in
several laboratories by both ISOL and in-flight techniques for
fusion studies. For the production of these beams, the ISOL technique has been
utilized at the pioneering facility of the Cyclotron Research Centre in
Louvain la Neuve (Belgium)\cite{ry02} and at the new facility SPIRAL in
GANIL (France)\cite{vi01}. The in-flight technique has been adopted at the
TWINSOL facility at the University of Notre Dame
(USA).\cite{be97,le99}

\subsubsection{The $^{6}$He+$^{209}$Bi system}

\begin{figure}[th]
\centerline{\psfig{file=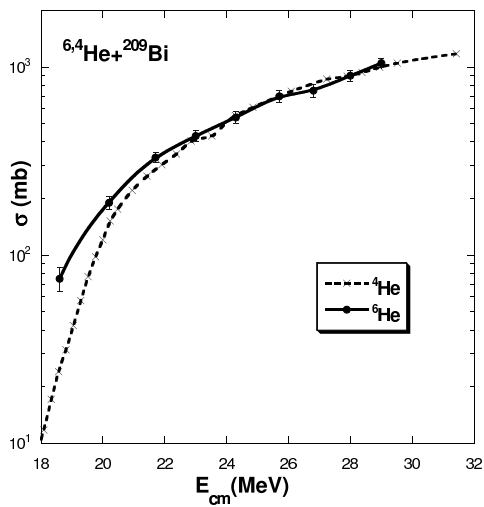,width=8cm}}
\caption{Fusion cross sections for the systems $^{4,6}$He+$^{209}$Bi. The 
$^{4}$He cross sections have been corrected for the different radius
and Coulomb barrier with respect to the $^{6}$He projectile.
\label{fg:he6bi209}}
\end{figure}

This system, with V$_{B}$ = 19.76 (20.33) MeV in the center of mass
(laboratory), has been studied extensively at Notre Dame. The $^{6}$He
beam is produced by the $^{9}$Be($^{7}$Li,$^{6}$He) proton transfer reaction
with the following characteristics: intensity $\sim 10^{5}$ pps, energy
ranging from 14 to 22 MeV, energy resolution $\sim$1.5 MeV, beam size
on target from 5 to 8 mm diameter.
The fusion cross section was evaluated by the sum of the
($^{6}$He,3n)$^{212}$At and ($^{6}$He,4n)$^{211}$At evaporation
channels.\cite{ko98} These data are essentially for complete fusion.
This cross section could be somewhat underestimated at the lowest energies
since the 2n channel, expected to be small, was not studied.
The possible incomplete fusion of $^{4}$He, produced by $^{6}$He
breakup, and the subsequent 1n emission to $^{212}$At was excluded by the
authors by reaction Q-value arguments. This should
eventually be verified experimentally. Within the present results, since there
is no incomplete fusion, the complete fusion measured has to be considered
as total fusion. This system, compared to $^{4}$He+$^{209}$Bi, shows
moderate enhancement only in the sub-barrier region as shown in
Fig.~\ref{fg:he6bi209}

\begin{figure}[th]
\centerline{\psfig{file=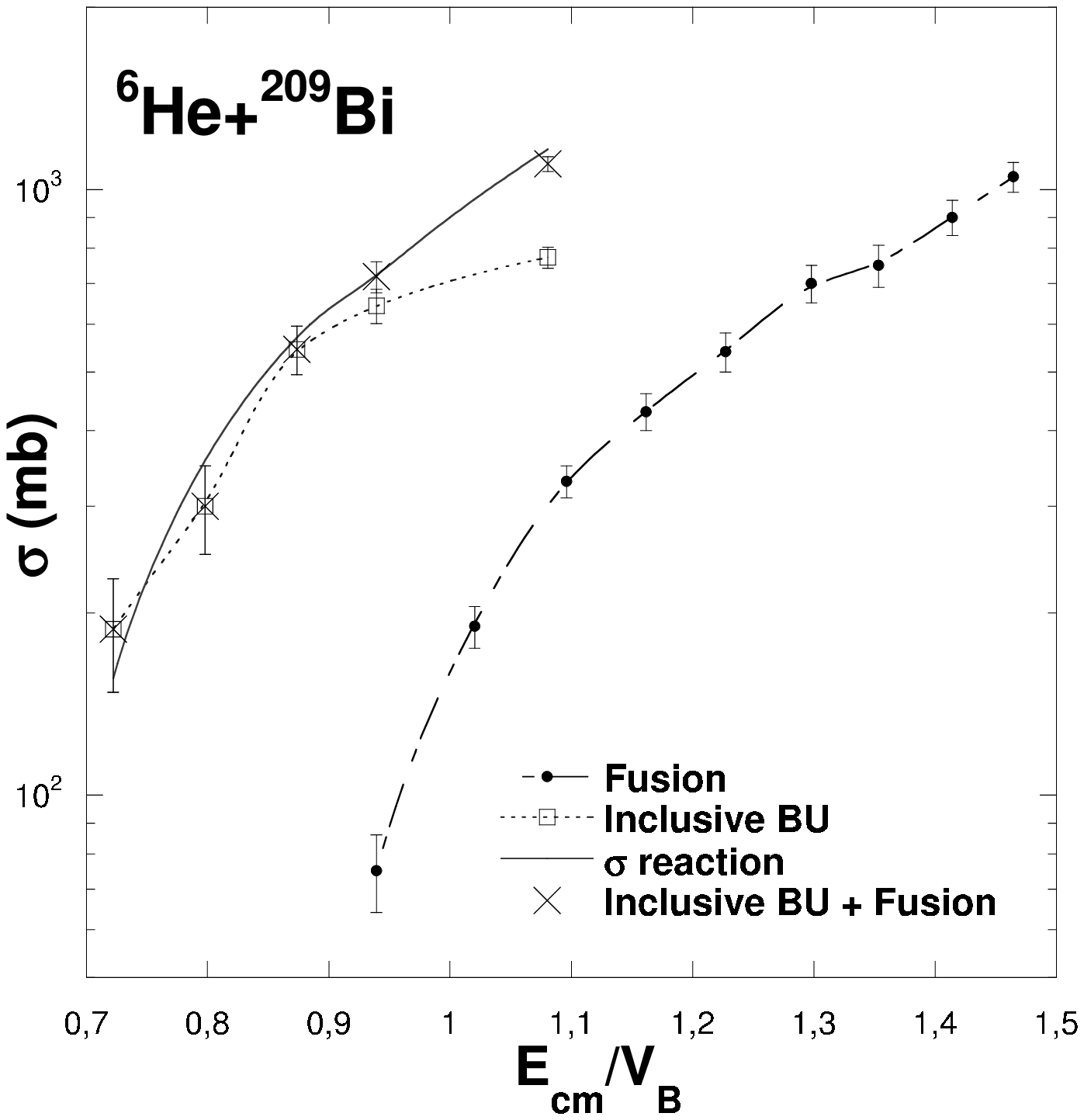,width=8cm}}
\caption{Cross sections for the following processes in the system
$^{6}$He+$^{209}$Bi: total reaction (solid curve), total fusion
(solid circles), inclusive $\alpha$ production or inclusive breakup
(open squares), total fusion+inclusive breakup (crosses). The total
fusion+inclusive breakup cross section exhaust the
total reaction cross section in the region where common data exist.
\label{fg:he6elas}}
\end{figure}

Above the barrier, the two cross sections, once corrected for the
different projectile radii and Coulomb barrier energies, are
essentially equal. This suggests that, in this case, the breakup effects have
negligible or no influence on the fusion process above the barrier.
The breakup process is quite relevant in this system. Indeed, the
so-called inclusive $\alpha$ production has been found very strong\cite{ag00}
as compared to fusion, particularly below the barrier. This $\alpha$ production
is most likely originating from the $^{6}$He breakup.
This seems to be the only relevant process in addition to the
fusion. As a matter of fact, breakup and fusion exhaust the total reaction
cross section inferred from the elastic scattering data\cite{ag01}, as shown
in Fig.~\ref{fg:he6elas}.

The detailed mechanism of the inclusive $\alpha$ production is not yet very
clear. These inclusive $\alpha$'s can originate by several processes, namely:
$\alpha$+n+n+$^{209}$Bi, $\alpha$+n+(n+$^{209}$Bi), and
$\alpha$+(2n+$^{209}$Bi). With
parenthesis we indicate the nucleus where one or two neutrons are trapped
into the target by the nuclear potential. These two
processes are of the transfer/stripping-breakup type
since they have a cross section much larger than
a conventional transfer process, usually well predicted by
direct reaction formalism like distorted-wave Born approximation (DWBA) or
coupled-channels.

Very recently $\alpha$-n coincidence measurements were undertaken
by the same Notre Dame group at 23 MeV, slightly above the
Coulomb barrier. In their first
experiment\cite{by04}, they studied the $^{209}$Bi($^{6}$He,$^{5}$He)$^{210}$Bi
reaction followed by $^{5}$He $\rightarrow$ $^{4}$He+n by measuring the
$\alpha$ particle and neutron in coincidence.
The neutrons were detected in liquid scintillators with
a relatively high threshold ($>$ 1 MeV). The experimental results give evidence
that the coincidence events, assigned to one neutron transfer/stripping-breakup,
account for approximately 20\% of the inclusive $\alpha$-particle yield which
has a cross section of 800 mb.

In their second experiment,\cite{de04} they studied
$^{209}$Bi($^{6}$He,$^{4}$He)$^{211}$Bi by detecting $\alpha$ particles at
and beyond the grazing angle in coincidence with neutrons evaporated from
$^{211}$Bi. The neutrons were detected in solid plastic scintillators with
a lower threshold ($>$0.3 MeV), but unfortunately with a high
background of around 45\% of the total neutron yield. In this case,
such coincidence yield accounts for about 55\% of
the inclusive $\alpha$-particle yield. The remaining 25\% yield, around 200 mb,
of the $\alpha$-particle inclusive events is assumed to originate from a
breakup process with all
$^{6}$He fragments, $\alpha$ and two neutrons, in the exit channel.

These results have similarity to the $^{6}$Li+$^{208}$Pb reaction\cite{si03}.
Lithium-6 with S$_{\alpha}$ = 1.47 MeV is the least bound stable nucleus.
In this system, strong inclusive $\alpha$ production was observed. The
breakup process, $\alpha$-d as well as $\alpha$-p coincidences, is
approximately one order of magnitude smaller with cross sections ranging from
65 mb to 110 mb around the barrier.

These results make the theoretical description of the $^{6}$He+$^{209}$Bi
reaction dynamics quite intriguing, since in addition to the
usual coupling to the target or projectile bound state
excitations and their consequent barrier distributions, one
has to consider also the coupling to the breakup channels, which
proceeds in most cases via continuum excitations since they lie above the
particle emission threshold.

\begin{figure}[th]
\centerline{\psfig{file=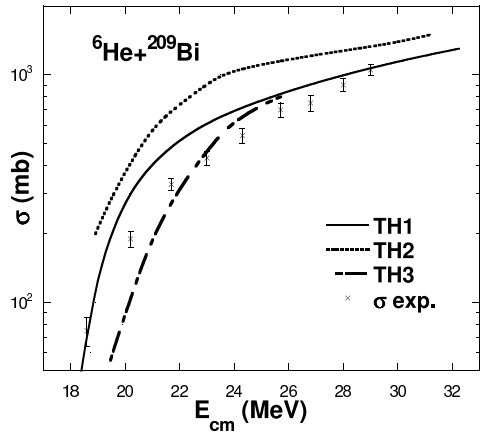,width=8cm}}
\caption{Theoretical calculations of the $^{6}$He+$^{209}$Bi fusion excitation
function. See text for details.
\label{fg:he6theor}}
\end{figure}

Several theoretical approaches have been
undertaken. A first attempt was done in Ref.~\cite{al02}. These calculations
(TH1) were done via the coupled-channel formalism using the {\tt ECIS}
code\cite{ecis}. The real potential was calculated with a double folding model
using the BMD3Y1 interaction\cite{bmd3y}. The imaginary potential had a small
radius in order to absorb all the flux penetrating the barrier simulating the
incoming wave boundary conditions. However the real potential had to be
reduced by a factor of 0.4 to reproduce the data. This
renormalization most likely takes care of the breakup process in a bulk way.
In Ref.~\cite{ru04}, two other approaches were followed. In the first one,
the calculations (TH2) were done with the
coupled discretized continuum channel (CDCC)
formalism and two slightly different ways of handling the potentials: a)
empirical cluster-target optical potentials, and b) short-ranged imaginary
parts. In these cases, the code {\tt FRESCO}\cite{fresco} was utilized.
The results of the CDCC b) approach,
a slightly better one in our opinion, are shown in the Fig.~\ref{fg:he6theor}.
In the second one, calculation (TH3) was done with the barrier penetration
model. In this case the potential was parameterized as
V$_{nucl}$~=~V$_{bare}$+V$_{pol}$, and the polarization potential was derived
from the previous CDCC approach.  In TH2 the breakup process is automatically
included in the CDCC formalism. In TH3 the breakup enters via the polarization
potential V$_{pol}$. See Ref.~\cite{ru04} for more details. The results of
these three different approaches are shown in the Fig.~\ref{fg:he6theor}.

From the inspection of Fig.~\ref{fg:he6theor} we see that TH3 and TH1 seem to
reproduce the experimental data better, however, not all the points.
These two
approaches are more phenomenological. TH2 is more fundamental since it is based
on
the CDCC formalism and reproduces well the elastic scattering as discussed in
Ref.~\cite{ru04} but fails as a good reproduction of the fusion. This last
problem
seems to be, for the moment, always present in this type of approach as we will
discuss later in connection with the $^{6}$He+$^{64}$Zn system. However,
it should be emphasized that the variation of these various theoretical
predictions is not necessarily bad. It
simply demonstrates the evolution and the present status of the theoretical
efforts and indicates that, most likely, additional work is
needed to evaluate the interaction of loosely bound nuclei at the barrier.

\subsubsection{The $^{6}$He+$^{238}$U system}
This system, with V$_{B}$ = 21.19 (21.73) MeV in the center of mass
(laboratory), has been measured twice at Louvain la Neuve.
The $^{6}$He beam was
produced via the ISOL\cite{ry02} method with
energy resolution $\sim$0.5\% and intensity $<5\times 10^{7}$ pps. The first
series of measurements was performed at eight energies from 14.6 to 28.7 MeV 
with four primary energies and four degraded energies
using mylar absorbers.\cite{tr00} The fusion process
was identified by the fission channel. The fission fragments were measured
with
a large solid angle detection system consisting of large area silicon surface
barrier detectors
arranged around the target in a box-like structure. In this first measurement,
the fission fragments originating from $^{6}$He induced fusion, $^{4}$He
induced fusion, and
1n or 2n transfer/stripping-breakup could not be distinguished, so
these results overestimate the total fusion cross section.

\begin{figure}[th]
\centerline{\psfig{file=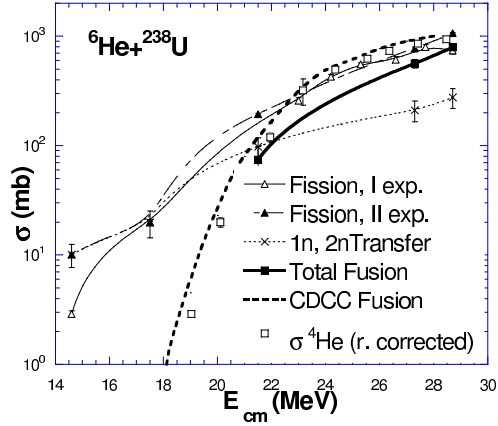,width=8cm}}
\caption{Cross sections measured in the system $^{6}$He+$^{238}$U. Total
fission from the first experiment (open triangles), from the second
experiment (filled triangles), yields of the $\alpha$-fission fragment
coincidences, {\it i.e.}, 1n or 2n transfer induced  fission (crosses),
and total fusion deduced from the
second experiment (filled squares). The relative theoretical fusion cross
section calculated using CDCC is shown (bold-dotted curve). The cross
sections for $^{4}$He induced fusion are also reported (open squares).
These last data include a small correction due to the
geometrical difference between the two projectile radii. The curves connecting
the various points are drawn to guide the eye. \label{fg:he6u238}}
\end{figure}

As discussed in several papers, the total fusion cross
section for $^{6}$He+$^{238}$U,
identified by all the fission events, is much larger, especially below the
barrier, than that for $^{4}$He+$^{238}$U by up to two orders of magnitude.
The
second experiment\cite{ra04} was set up in order to also measure the
$\alpha$-fission fragment coincidences. In this way it was possible to
distinguish the fission events originating from
total fusion (not in coincidence with $\alpha$ particles) from fission events
originating from 1n or 2n transfer/stripping-breakup.
Figure~\ref{fg:he6u238} presents the results of the
two experiments as well as the cross sections with $^{4}$He beam for
comparison. First of all, it should be noted that at some energies the
cross sections measured in the two runs differ well beyond the statistical
errors; see, for example, the lowest energy point. In our opinion, this could
originate
from systematic errors due to the low beam intensity and the related problems
of correctly identifying the reaction channel. The most interesting result is,
however, that fission below the barrier is only in coincidence with $\alpha$
particles and consequently it originates by transfer/stripping-breakup. Such a
process should not be included in incomplete fusion. This
result is very similar to that of the $^{6}$He+$^{209}$Bi reaction.

It is rather surprising that, at this level of accuracy, total fusion (only
three points) is not at all reproduced by the CDCC calculations based on
the work in Ref.~\cite{ru03} and that the $^{6}$He induced total
fusion is smaller than $^{4}$He induced fusion
{\it i.e.}, fusion is hindered not enhanced. However,
the $^{4}$He induced fusion at the highest energies originate from other data
as stated also in Ref.\cite{al02}. These points
should be investigated further, experimentally and theoretically.

\subsubsection{The $^{6}$He+$^{64}$Zn system}

This experiment was done at the Cyclotron Research Centre in Louvain la Neuve
(Belgium) using the two coupled cyclotrons via the ISOL method\cite{ry02} with
$\sim 3\times 10^{6}$ pps beam intensity. Four beam energies were explored in
the range of 10 to 13.6 MeV (V$_{B}$ = 9.5 (10.4) MeV in the center of mass
(laboratory) system). The $^{6}$He primary beam had the
energy fixed at the maximum value and the lower energy points were obtained by
a suitable number of niobium  absorbers $\sim$ 3 mg/cm$^{2}$ thick.
The experimental set up was arranged in order to measure
a) the X-rays emitted following the electron capture decay of various
evaporation residues offline, and
b) the $^{6}$He scattering events and the inclusive $\alpha$-particle
channels originating most likely from transfer/stripping-breakup. For
this second part, the large area segmented detector LEDA was
utilized.\cite{da00} For comparison, the $\alpha$+$^{64}$Zn system was
measured with the same setup at three energies overlapping those
of the $^{6}$He+$^{64}$Zn system.

The main results\cite{di04} are summarized here.
A strong population of the $^{65}$Zn (= n+target) nucleus was observed.
The cross section is much larger than the $\alpha$n evaporation yield
predicted by the statistical model code {\tt CASCADE}\cite{cascade}.
The cross section for the 2n(producing $^{68}$Ge), pn($^{68}$Ga),
p2n($^{67}$Ga), and $\alpha$n ($^{65}$Zn)
channels were measured. Their sum is
essentially the total fusion cross section if the $\alpha$n contribution is
excluded, since the $^{4}$He+target incomplete fusion produces, most likely,
the same X-ray emitters $^{67}$Ge (1n evaporation) and $^{67}$Ga (1p
evaporation) as the $^{6}$He+target complete fusion.
This total fusion could be slightly underestimated since there
could be some missing channels, like the 3n(1n) emission from
$^{6}$He($^{4}$He) capture, which, however, are
expected
to contribute a negligible amount. Still, the comparison with the
$^{4}$He+$^{64}$Zn system measured in the same way is quite meaningful.

Once the measured $\alpha$n contribution is replaced by that calculated
by {\tt CASCADE}, and trivial
geometric corrections are included there is no difference between $^{6}$He
and $^{4}$He+$^{64}$Zn cross sections in the few points measured, {\it i.e.},
neither enhancement nor hindrance.
The reaction cross sections extracted from the elastic scattering data
confirm the existence of strong reaction channel(s) in addition to fusion.
The potential extracted from the optical
model fit to the elastic scattering data, with $r_{0}$ = 1.2 fm, does not
reproduce the fusion cross sections
with the {\tt CCFULL}\cite{ccfull} calculations. For a good
agreement, a smaller value of $r_{0}$ = 1.0 fm has to be used in the
calculations. But such a choice is difficult to justify on the basis
of measured nuclear radii and related systematics, unless $r_{0}$ is treated
simply as an adjustable parameter to reproduce experimental data.
The fact that the potential
extracted from the elastic scattering data does not reproduce the fusion data
is not unexpected. A similar conclusion was already pointed out; see in
particular Ref.~\cite{ne04}. The interpretation of this is that the elastic
scattering samples a different portion of the potential, more peripheral than
the one involved in fusion. Therefore, this potential does not necessarily
apply to the fusion process.

\subsubsection{The $^{6,8}$He+$^{63,65}$Cu and $^{6}$He+$^{188,190,192}$Os}

These measurements\cite{na04} were performed at GANIL/SPIRAL where the
radioactive beams are produced via the ISOL method.\cite{vi01} The $^{6}$He
($^{8}$He) beam had an intensity of 10$^{7}$ ($7\times 10^{4}$) pps, a beam
spot on the target of 5 (8) mm in diameter, and a good resolution of 0.1\% in
both cases. The following energies were utilized: $^{6}$He
+$^{65}$Cu 19.5 and 30 MeV (V$_{B}$ = 9.7 MeV), $^{6}$He+$^{63}$Cu 30 MeV,
$^{8}$He+$^{63}$Cu 27 MeV, and $^{6}$He+Os targets 30 MeV (V$_{B}$ = 21 MeV).
The
detector array consisted of 8 $\gamma$-ray clover detectors from EXOGAM located
10.5 cm from the target and an annular segmented SSD
(16 rings x 16 segments) for charged particle detection positioned around 
0 degrees at $\sim$3.5 cm from the target. This way the system could detect
in beam $\gamma$ rays, forward emitted charged particles (CP), and CP-$\gamma$
coincidences.

With the Cu targets the following data were obtained: the cross sections for 
production of various evaporation residues and/or transfer products (from the
characteristic $\gamma$ rays of the nuclei populated), Q-value
spectra, and elastic scattering angular distributions which yielded total
reaction cross sections. It should be pointed out that this is the first time
that fusion cross
sections induced by RIBs could be measured via in beam $\gamma$-ray techniques.

In the $^{6}$He+$^{65}$Cu system measured at
two energies, it was observed that
the population of $^{66}$Cu (=1$n$+$^{65}$Cu) is $\sim$10 times stronger than
the $\alpha$n evaporation channel predicted by statistical model calculations.
The analysis of the Q-value spectra indicates that $^{66}$Cu is populated
mainly from the 1$n$ evaporation from $^{67}$Cu produced by 2n+$^{65}$Cu.
The fusion cross section, most likely complete fusion, plus this
strong transfer/stripping-breakup cross section constitute the largest amount,
85\%,
of the total reaction cross section deduced from the elastic scattering data.
The remaining 15\% is, most likely, the exclusive breakup cross section
with all fragments in the exit channel.
The $^{4}$He+$^{63,65}$Cu systems were investigated in parallel at the Bombay
14UD BARC-TIFR tandem accelerator (India) via a similar in beam gamma-ray
technique. In this case the strong transfer/stripping-breakup channel was
absent.

The data collected with the Os targets have, in some cases, lower statistics,
and for the $^{190}$Os target the absolute cross sections could not be
determined. Nonetheless, the Os targets results are consistent with the
Cu target results.

\subsubsection{Comments on the $^{6}$He induced reactions}

For the total fusion measured with $^{209}$Bi, $^{238}$U, and $^{64}$Zn
targets, only the $^{6}$He+$^{209}$Bi system shows enhancement with respect to
the $^{4}$He total fusion, but only for the two lowest energy points. This is,
for the moment, attributed to the coupling to breakup
excitations, as predicted by theories. A remeasurement of these cross sections
with higher statistics would be desirable.
The really new and strong effects originating from the breakup process
are the inclusive and exclusive alpha particle yields. The
strongest channels are assigned to the formation of the systems (n+target)
and (2n+target). These are most likely not the conventional transfer
processes,
usually with moderate cross sections, but processes where unbound, highly
excited states are formed in a sort of compound nucleus.

\begin{figure}[th]
\centerline{\psfig{file=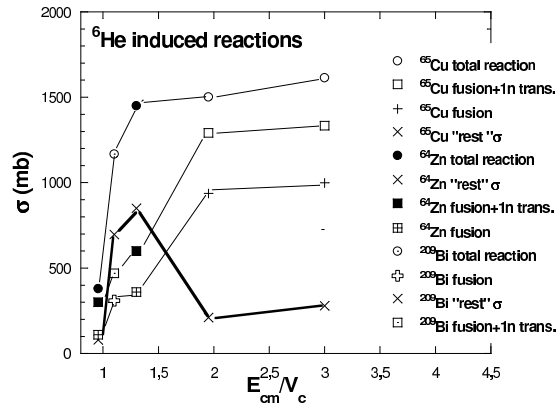,width=8cm}}
\caption{Experimental cross sections for the following processes induced by
$^{6}$He on $^{65}$Cu, $^{64}$Zn  and $^{209}$Bi targets: total fusion, total
fusion + 1n transfer/stripping-breakup, total reaction, and
residual cross sections (labelled by ``rest'' in the figure). The residual
cross section is the difference between the total reaction and the total
fusion + 1n transfer/stripping-breakup cross sections.
\label{fg:he6comm}}
\end{figure}

In order to search for possible common features in the $^{6}$He induced
reactions, we have plotted in Fig.~\ref{fg:he6comm} the following
data, total reaction cross section, fusion+1n transfer/stripping-breakup
cross section, total
fusion cross section, and residual cross section as a function of
E$_{cm}$/V$_{B}$ with V$_{B}$ computed with $r_{0}$ = 1.56 fm. The residual
cross section is the difference between the total reaction and
the fusion+1n transfer/stripping-breakup cross sections.
It should account for all the processes not included in
the fusion and in the 1n transfer/stripping-breakup which are essentially the
2n transfer/stripping-breakup and the
exclusive breakup with all $^{6}$He fragments in the exit channel.

Fusion and fusion+1n transfer/stripping-breakup cross sections increase
continuously and
smoothly with E$_{cm}$/V$_{B}$. This is essentially the increase of
fusion cross section with rising bombarding energy. So all three
systems seem to behave in a similar way in this aspect.

The scenario is somehow different for the total reaction cross sections and the
related residual cross sections. The residual cross sections, according to the
Zn and Bi data, have a maximum around the barrier. The actual height of this
maximum, with respect to the Cu target data, is in part
questionable since the authors state
that the 1n transfer/stripping-breakup events could also originate from
1n evaporation from the
system 2n+target.\cite{na04} A possible interpretation of this maximum is
that around
the barrier the breakup phenomena are the strongest. This behavior is peculiar
because it is different from the other two cross sections and deserves further
investigation since this is the first time that it is observed. Anyhow, since
these
various cross sections were measured by different experimental methods
they should be confirmed by check experiments.

\subsection{$^{11,10}$Be beams}

\begin{figure}[th]
\centerline{\psfig{file=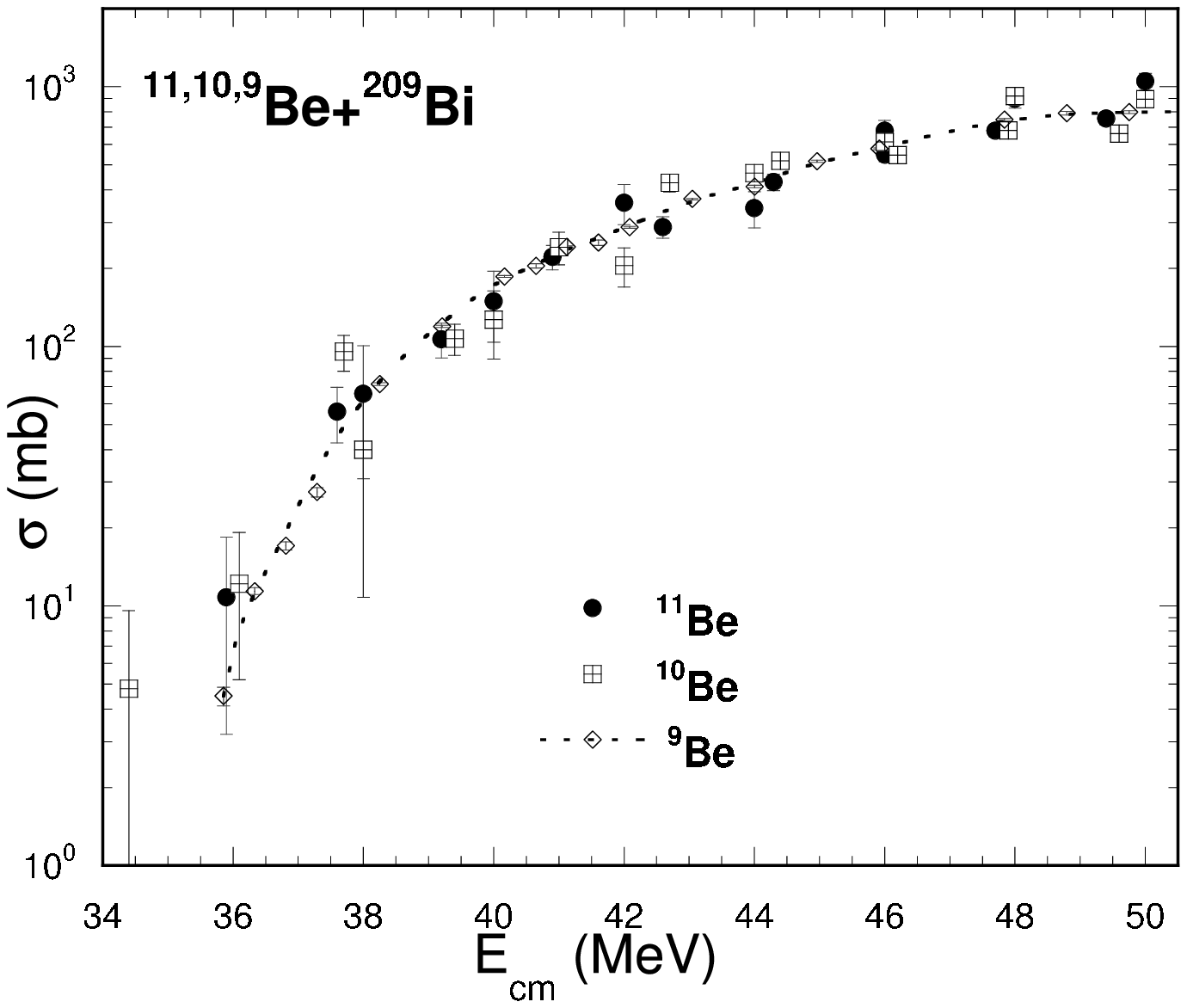,width=8cm}}
\centerline{\psfig{file=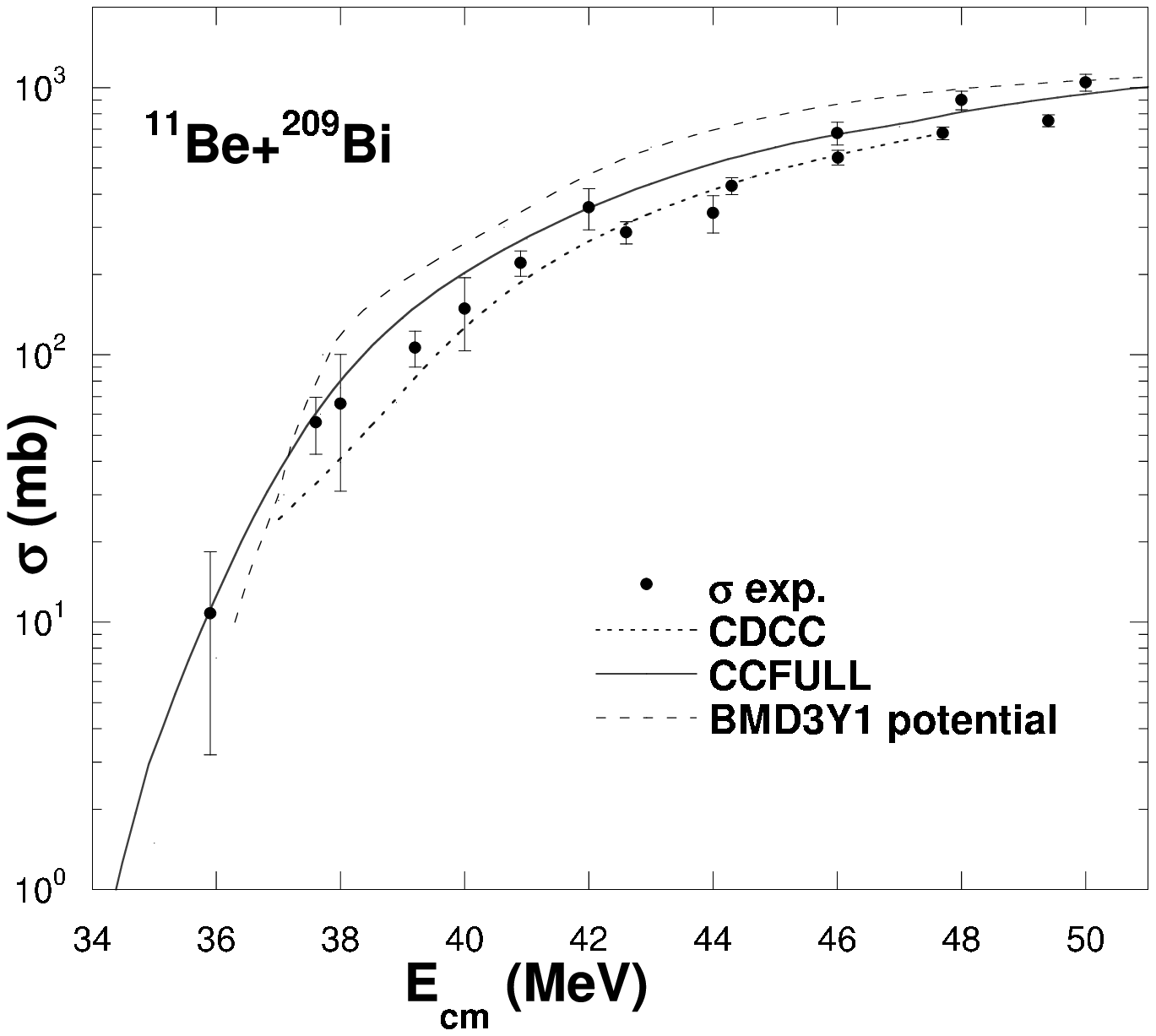,width=8cm}}
\caption{Total fusion cross sections for the systems $^{9,10,11}$Be+$^{209}$Bi
(top) and comparison between theories and experimental cross sections for the
$^{11}$Be+$^{209}$Bi system (bottom).
\label{fg:be11bi209}}
\end{figure}

These experiments were done at RIKEN in the RIBS beam line.\cite{ku92} The
$^{11,10}$Be beams were produced in-flight by fragmentation of a
100 MeV/nucleon $^{13}$C primary beam on a thick beryllium production
target. The Be beams had energy $\sim$10 times larger than what is
needed for fusion measurements at Coulomb barrier energies. The Be beams were,
therefore,
heavily degraded producing the final beam with a large energy spread,
35 to 55 MeV, and poor emittance, $\sim$5 cm diameter at a $^{209}$Bi target.
The Be
beam intensities finally achieved were $>$10$^{5}$ pps. The energy of the beam
particles producing fusion events were tagged event-by-event via
the time-of-flight over a flight path of $\sim$6 m. The detection system
consisted of large area SSDs each with an active area of 5x5 cm$^{2}$ arranged
in a compact box-like
structure as close as possible to a multitarget setup. The fusion events were
identified by $\alpha$ particles, with characteristic lifetimes and
energies, emitted in the decay of the various evaporation residues populated
after neutron evaporation from the compound nucleus and by the fission
fragments in coincidence in two opposite SSD of the box.\cite{si98,si04} The
fusion cross sections were identified as the sum of the fission and 4n+5n
(3n+4n) evaporation channels for $^{11}$Be ($^{10}$Be). These fusion cross
section could be slightly underestimated below the barrier since the 3n
(2n) channel for $^{11}$Be ($^{10}$Be) could not be measured due to its too
long lifetime, T$_{1/2}$ = 16 $\mu$s, which resulted in high random
rates. Moreover, the evaporation channels with
at least one charged particle, like p$x$n, were not identified, since
they are expected to be negligible. Even with these
limitations, the comparison of $^{11}$Be and $^{10}$Be fusion is quite
meaningful since the data were measured and analyzed in the same way.
For $^{10}$Be, the complete fusion cross section is taken as the total
fusion because no breakup processes are realistically expected since $^{10}$Be
with
S$_{n}$ = 6.8 MeV is tightly bound. For $^{11}$Be, mainly total fusion was
measured since the
incomplete fusion of $^{10}$Be, from $^{11}$Be breakup, was estimated in a
previous
experiment\cite{yo96} to be $<$30\% of the $^{11}$Be complete fusion.
These cross
sections came from two independent measurements done some years apart.
The statistics of the data is essentially limited by
the low intensity of the radioactive beams. This causes a scattering of the
various results beyond the statistical errors as discussed in detail in
Ref.~\cite{si98,si04} The results are compared in the top panel of
Fig.~\ref{fg:be11bi209} as well as with the stable $^{9}$Be beam.\cite{si99}

Within the statistical accuracy of the data, the three cross sections look
similar. This is astonishing since $^{10}$Be is tightly bound and both $^{9}$Be
and $^{11}$Be are loosely bound. In addition, the latter nucleus has a well
established neutron halo structure while $^{10}$Be is well bound like most of
the stable nuclei. Three theoretical approaches have been
followed to describe these results:
\begin{itemize}
\item[a)] {\tt CCFULL}:\cite{si98,si04} The main coupling
considered was the
excitation of collective, rotational-like structure, with no breakup coupling.
\item[b)] CDCC using the code {\tt FRESCO}\cite{fresco}:\cite{si98,si04}
Only the two-body breakup
coupling to discretized continuum states was calculated. These
last calculations need several days of CPU time on modern fast PCs.
\item[c)] coupled channel calculations using the {\tt ECIS} code\cite{al02} as
already mentioned in the discussion of the $^{6}$He+$^{209}$Bi system.
\end{itemize}

The bottom panel of Fig.~\ref{fg:be11bi209} shows the results of the first two
approaches which are similar. But, if we consider also the other two
beams $^{10}$Be and $^{9}$Be, the CDCC approach, particularly in the $^{9}$Be
case\cite{si98,si04} not shown in Fig.~\ref{fg:be11bi209}, underpredicts
the cross sections. Apparently, the collective excitations, better handled by
the {\tt CCFULL} code have more strength than the breakup mode.
The third approach looks, on average, worse, since it overestimates the cross
section at most of the energies. But, as already pointed out, the breakup
coupling is
included in an indirect way via renormalization of the nuclear potential.
The large collective excitation strength versus the breakup strength
seems to appear
also in the scattering of loosely bound $^{17}$F
(S$_{p}$ = 0.600 MeV) compared with the well bound $^{19}$F (with
possible collective structure), as later discussed in this review (Sect. 
{\bf 4.4}).
In the case of $^{19}$F, at energies around the Coulomb barrier, the
reaction cross section is larger than $^{17}$F, most likely, due to
a rotational-like level
structure which may be excited more easily than $^{17}$F breakup.

For completeness, we should mention that the first attempt at GANIL to measure
the $^{11}$Be+$^{238}$U system\cite{fe99} leads to the successful measurement
of $^{6}$He+$^{238}$U with the same technique.\cite{tr00}
The statistics were
very limited so that no conclusion can be drawn from the data.

\subsection{$^{11}$C beam}

This experiment was performed at the 88-inch cyclotron at
Lawrence Berkeley National Laboratory. The system studied was
$^{11}$C+$^{197}$Au.\cite{jo00} This experiment is worth mentioning
mainly for the technique used for the production of this RIB. The
system utilized was BEARS (Berkeley Experiment with Accelerated
Radioactive Species) which was based on two coupled cyclotrons. Carbon-11,
$t_{1/2}$ = 20 min., was produced via the $^{14}$N(p,$\alpha$)
reaction in a 20 atm nitrogen gas target using the cyclotron of the Biomedical
Isotope Facility, located 350 m away from the 88-inch cyclotron, and then
transported by a dedicated transfer line to this latter cyclotron for further
acceleration. This system provided a continuous $^{11}$C beam with a
remarkable intensity of 1-2$\times 10^{8}$ pps on target. This experiment
measured the excitation functions for the population of the evaporation
residues, following $x$n evaporation from the compound nucleus. These cross
sections were deduced by the yield of the $\alpha$ particles emitted from
the ground state decay of the various At nuclei produced. The excitation
functions were measured in the range of 66 to 110 MeV. The results compared
with the stable $^{12}$C beam are in agreement with standard
evaporation model predictions by the {\tt HIVAP} code\cite{hivap}.

\subsection{$^{17}$F beam}
%Sections, sub-sections and sub-subsections are numbered in
%Arabic.  Use double spacing before all section headings, and
%single spacing after section headings. Flush left all paragraphs
%that follow after section headings.

In contrast to most of the experiments discussed in the previous subsections
where the interplay of neutron breakup and fusion is investigated, this
subsection explores the influence of proton breakup on fusion.
Fluorine-17 is a proton drip-line nucleus with a ground state binding energy
of 0.600 MeV. Its first excited state is only bound by 0.105 MeV with an 
extended r.m.s. radius and is considered a halo state.\cite{mo97}
If the $^{17}$F nucleus is
excited to its first excited state before fusion, the fusion cross section
is expected to increase because of the lowered barrier for a larger radius.
In addition, if $^{17}$F breaks up prior to fusion, the core nucleus $^{16}$O
has a lower Z and a lower Coulomb barrier. The incomplete fusion is expected to
be
enhanced, too. On the other hand, as discussed in the previous section, breakup
removes the $^{17}$F flux available for fusion resulting in fusion suppression.
Measurements of fusion of $^{17}$F and $^{208}$Pb was performed
with the radioactive $^{17}$F produced by an in-flight method using the
$p$($^{17}$O,$^{17}$F)$n$ reaction at ATLAS.\cite{re98}
The $^{17}$F beam intensity was between
1 and 2$\times 10^{5}$ pps with an energy resolution of about 2.5\%.
There was a significant $^{17}$O isobar contamination in the beam with the same
magnetic rigidity. The ratio of $^{17}$F to
$^{17}$O varied with beam energies but was usually around 1. Since the
energy of the $^{17}$O contaminant was approximately 20\% lower than that of
$^{17}$F,
the fusion of $^{17}$O with the target was estimated to contribute less
than 3\% to the total fusion cross section. The compound nucleus decays by
fission, therefore, the fusion cross section was
identified with the fission cross section. The fission fragments were
detected in coincidence by four large area Si detectors. It is noted
that incomplete fusion,
capture of $^{16}$O by the target following
$^{17}$F $\rightarrow$ $^{16}$O+$p$
breakup, was not excluded from this measurement, so the data give the total
fusion yield (CF+ICF).

The measured fusion-fission excitation function is compared to that of
$^{19}$F+$^{208}$Pb and $^{16}$O+$^{208}$Pb. No fusion enhancement was observed
for $^{17}$F induced fusion with respect to the stable
$^{19}$F and $^{16}$O induced fusion on the same target. At the lowest energy
of the measurement, the fusion-fission cross section is suppressed by a factor
of 4.

\begin{figure}[th]
\centerline{\psfig{file=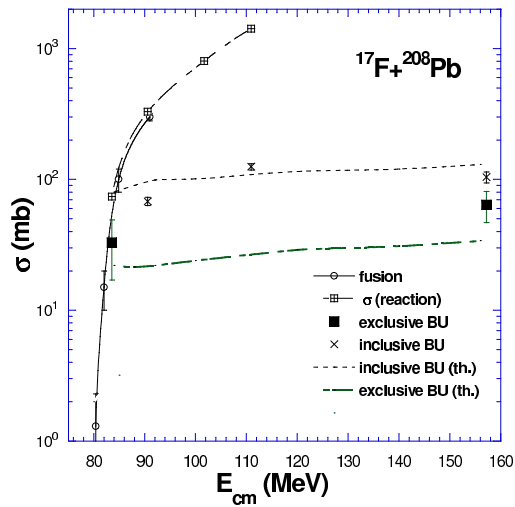,width=8cm}}
\caption{The excitation function of $^{17}$F+$^{208}$Pb
fusion\protect\cite{re98}
is shown by the open circles, inclusive breakup\protect\cite{li03a}
by the crosses, and exclusive breakup\protect\cite{ro04} by
the filled squares. The reaction cross sections obtained from the optical
model analysis of elastic scattering data are shown by the thin dash-dotted
curve.
The dashed and dash-dotted curves are for stripping and diffraction breakup,
respectively.\protect\cite{li03a} \label{fg:f17pb}}
\end{figure}

The breakup of $^{17}$F may be a factor in influencing fusion. Measurements of
$^{17}$F breakup by scattering on a $^{208}$Pb target were performed at
energies
above the Coulomb barrier (98, 120, and 170 MeV) at HRIBF (Oak Ridge National
Laboratory)\cite{li00,li02,li03a} and below the barrier (90 MeV) at
ATLAS\cite{ro04}. The HRIBF produced $^{17}$F by the ISOL technique.  
The intensity was 2$\times 10^{5}$ pps for the first
measurement and increased to $10^{7}$ pps for subsequent measurements
a year later.
Since the secondary beams were accelerated by a tandem accelerator,
the beam quality (energy resolution and beam spot) was as good as for
stable beams. The $^{17}$O isobar contaminants were removed by selecting the
9$^{+}$ charge state using an analyzing magnet. Two
modes of breakup, diffraction (exclusive, two-body) and stripping(inclusive,
one-body), were measured at 170 MeV. The
diffraction breakup which has $^{16}$O and $p$ in the exit channel, was found
to be a factor of four smaller than stripping breakup (only $^{16}$O in the
exit channel). This observation agrees with predictions by Esbensen.\cite{li02}
At energies near the barrier, only the stripping breakup was measured. The
energy dependence of the breakup is presented along with that of fusion in Fig.
\ref{fg:f17pb}. The stripping breakup cross sections are smaller than the
fusion cross sections by factors of 4 to 10 and the diffraction breakup
is four times smaller than the stripping breakup. At energies below the
barrier, the diffraction breakup was
studied at backward scattering angles.\cite{ro04}
The angle integrated breakup cross
section agrees with the theoretical prediction and is very small. 
It is concluded that breakup of $^{17}$F is weak and it has no noticeable
influence on fusion.

The elastic scattering of $^{17}$F+$^{208}$Pb near the barrier was
analyzed by an optical model.\cite{li03a,ro04}
Due to the thick targets used in these experiments,
the energy resolution was poor. The excitation
of $^{17}$F to its first excited state as well as other quasielastic
scattering, resulting in a small energy and/or mass change, could not be
separated from the pure elastic scattering. The shape of the angular
distribution for $^{17}$F+$^{208}$Pb elastic scattering is similar to that of
$^{16,17}$O+$^{208}$Pb, particularly at backward angles. These differential
cross sections are much smaller than in the $^{19}$F+$^{208}$Pb system. This
indicates a smaller absorption cross section in the
$^{19}$F+$^{208}$Pb. The total reaction
cross sections obtained from the optical model analysis are slightly
larger than the fusion cross sections, as shown in Fig. \ref{fg:f17pb}.
The small difference between the total reaction and fusion cross sections
suggests that other reaction channels such as breakup are weak and
fusion enhancement is not expected. It was speculated by Romoli {\it et al.}
that the differences in the elastic scattering between $^{17}$F and $^{19}$F
may arise from the
large collectivity of $^{19}$F
where the ground state is strongly coupled to its first excited
state.\cite{ro04} As a result, the excitation probability of $^{19}$F could
be significantly larger than the breakup probability of $^{17}$F. Furthermore,
calculations show that the excitation of $^{17}$F to the first excited
state is about 30\% larger than the diffraction breakup.\cite{re98} Therefore,
fusion following excitation to the halo state could be too small to be seen.

A large suppression of complete fusion was observed in $^{9}$Be+$^{208}$Pb
at energies
above the barrier.\cite{da99} The sum of evaporation residue and fission cross
sections accounts for 68\% of the expected fusion cross section. In their
measurements, the capture of a charged fragment from the breakup of $^{9}$Be
by the target (incomplete fusion) was observed. A similar situation occurs also
in $^{9}$Be+$^{209}$Bi.\cite{si99} The complete and incomplete
fusion make up the expected fusion cross sections. In the $^{17}$F experiments,
elastic scattering and breakup reaction data already exist. Further
measurements of complete and incomplete fusion separately in
$^{17}$F+$^{208}$Pb would be useful for a complete theoretical description of
breakup and fusion.

It should
be pointed out that dynamical polarization is predicted to be present when a
loosely bound nucleus approaches a heavy nucleus with large Coulomb
fields.\cite{es96,es02}
Moreover, the dynamical polarization of a proton loosely bound nucleus will
push the valence proton to the far side and get shielded from the core nucleus.
This will
reduce the breakup probability. In contrast, a neutron loosely bound nucleus
tends to be polarized in such a way that the valence neutron lies between the
core nucleus and the target which leads to large breakup cross sections.
This may be the reason
for low breakup cross sections for $^{17}$F but large breakup cross sections
for $^{6}$He.

\section{Heavy-ion reactions}
In the following subsections, fusion induced by 
medium-mass neutron-rich radioactive nuclei is presented. This topic is of
particular interests for the potential
use of neutron-rich RIBs for producing heavy elements. It is noted that
for the following systems involving neutron-rich RIBs the excitation functions
will be compared in reduced coordinates using the Bass model fusion
barrier\cite{ba74}, and $\pi$R$^{2}$ with
R=1.2(A$_{p}^{1/3}$+A$_{t}^{1/3}$),
where A$_{p}$ and A$_{t}$ are the mass of the projectile and target,
respectively, to normalize the center-of-mass reaction energy and the cross
section. This is different from the procedures described in the light-ion
section because the medium-mass radioactive nuclei used in experiments so far,
are not very extended or loosely bound.

\subsection{$^{29,31}$Al beams}
The radioactive Al beams were produced by fragmenting 90 MeV/nucleon $^{40}$Ar
at RIKEN. Thick Al degraders were used to decrease the secondary Al beams to
the appropriate region of energy. This resulted in a very large energy
spread in the beams. It was necessary to use event-by-event time-of-flight
measurements to define the energy of each beam particle. They were also used
in the data analysis for setting gates to remove events originating from
contaminant induced reactions. The beam intensity
was 1$\times 10^{5}$ and 3$\times 10^{4}$ pps for $^{29}$Al and $^{31}$Al,
respectively. To avoid changing beam energies, a
stack of ten mylar backed $^{197}$Au targets was placed in series along the
beam direction. The excitation function was obtained in one run.\cite{wa01}

The compound nuclei formed in these reactions decay by fission. Therefore,
the measured fission excitation function was taken as the fusion excitation
function. The fission fragments were identified by two pairs of multiwire
proportional counters (MWPC) placed
on each side of the target stack. Valid fission events required a simultaneous
detection of both fragments with the MWPCs. The target from which the fission
fragments originated was identified by reconstructing the tracks of the
fragments. The cross
sections were scaled and checked with separate measurements using stable
$^{27}$Al beams.

\begin{figure}[th]
\centerline{\psfig{file=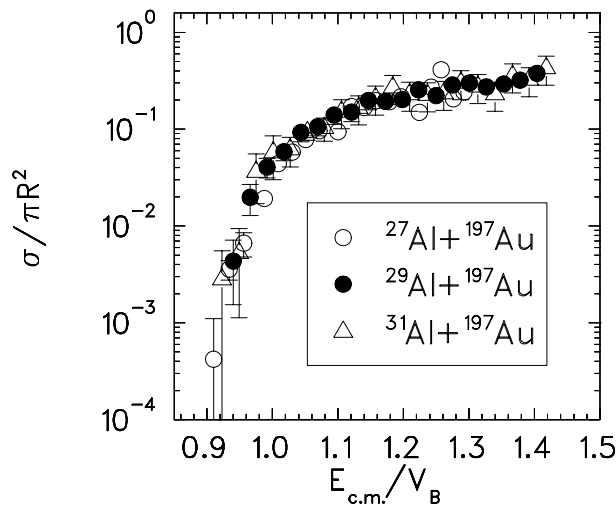,width=8cm}}
\caption{Comparison of fusion excitation functions for
$^{27,29,31}$Al+$^{197}$Au in reduced coordinates. The
open circles are for $^{27}$Al induced fusion, the closed circles are for
$^{29}$Al, and the open triangles are for $^{31}$Al.
\label{fg:al197au}}
\end{figure}

All three excitation functions, $^{27,29,31}$Al+$^{197}$Au, exhibit large
enhancement near and below the barrier with respect to barrier penetration
model predictions.\cite{wa01}
The enhancement cannot be accounted for by coupled-channel
calculations including excitation of the projectile and target, and the static
deformation of the target.
Nevertheless, when the excitation functions are plotted
in reduced coordinates which factor out the differences in nuclear sizes and
barrier heights, they overlap each other, as shown in Fig.~\ref{fg:al197au}. 
In order to avoid the influence from coupling to the intrinsic degrees of
freedom at energies near the barrier, the cross sections from high energy
measurements were fitted with $\sigma=\pi R^{2}(1-V_{b}/E)$
to extract the barrier height $V_{b}$ and
barrier radius R$_{b}$. It was found that the barrier height for $^{29}$Al and
$^{31}$Al induced fusion with respect to $^{27}$Al was reduced by 3.4 and 4.5
MeV, and the barrier radius was decreased by 0.1 and 0.2 fm, respectively.
In contrast, the Bass model\cite{ba74} predicts that the reduction of barrier
height and
the decrease of barrier radius is 1.2 MeV and 0.2 fm for $^{29}$Al induced
fusion as compared to $^{27}$Al induced fusion. The larger reduction in
measured barrier
height compared to the Bass model prediction may be due to the large
deformation in $^{27}$Al.

Watanabe {\it et al.} fitted the excitation functions with
Stelson's model assuming a flat distribution of barriers.\cite{st90} The
threshold barrier for neutron flow correlates with the binding energy of the
participants. The neutron binding energies are 13.1, 9.4, 7.2, and 8.1 MeV for
$^{27,29,31}$Al and $^{197}$Au, respectively. The neutrons are likely to
flow from Au to $^{27,29}$Al and from $^{31}$Al to Au. Although the
threshold barrier height and the direction of flow are different in the three
systems, the barrier thickness and the sub-barrier enhancement are very
similar, which is surprising but also puzzling.\cite{wa01} 

It is noted that the Q-value for two and four-neutron
pickup in $^{27}$Al and two-neutron stripping in $^{31}$Al are positive. All
the other neutron transfer reactions (up to four neutrons) have
negative Q-values. For this reason, the fusion is not likely to be enhanced by
sequential neutron transfer. This may distinguish neutron flow from neutron
transfer as a mechanism that can enhance the sub-barrier fusion rate.
On the other hand,
the Q-values for proton stripping reactions become more positive as the Al
isotopes become more neutron-rich . For instance, the Q-value is as large as
16.3 MeV for $^{197}$Au($^{31}$Al,$^{34}$S)$^{194}$Os.
The coupling of proton transfer reactions, which was not considered in the
previous calculations, may be important in neutron-rich radioactive nuclei
induced fusion.\cite{he87}

\subsection{$^{38}$S beam}
This experiment compares the fusion excitation function of radioactive
$^{38}$S on $^{181}$Ta to that of stable $^{32}$S on the same
target.\cite{zy97,zy01} Since
neither the sulfur isotopes nor the $^{181}$Ta nucleus is magic,
any effects due to  the shell closure can be realistically excluded.
The $^{38}$S beams
were produced by projectile fragmentation of a 40 MeV/nucleon $^{40}$Ar beam
on a $^{9}$Be target at NSCL. The energy of the
$^{38}$S beam delivered to the experiment was 8 MeV/nucleon, therefore,
beam degradation was necessary.
This was achieved by using Al foils of various thickness mounted at the
entrance of the scattering chamber to degrade the beam to the desired
energies, 161.2 to 254.0 MeV. An energy spread of 2.5 MeV FWHM was observed
as a result of the
degradation process. The beam intensities on target ranged from
2$\times 10^{3}$ to 1$\times 10^{4}$
pps with 85\% to 90\% $^{38}$S. 
Two sets of timing channel plates and parallel plate avalanche counters (PPAC)
were used for event-by-event beam tracking and
time-of-flight measurement. According to a statistical model prediction,
more than 99\% of the compound nucleus formed in this reaction decays
by fission. The measured fission excitation function was taken as the fusion
excitation function. The fission fragments were detected in an array
of PPACs and Si detectors. A valid event was defined by fragment-fragment
coincidence gated by time-of-flight associated with the $^{38}$S beam.
The fission
cross section was normalized to the Rutherford scattering measured by a
forward angle Si detector.

Quasifission usually complicates fusion-fission measurements in heavy
systems. In quasifission, the interacting nuclei are captured inside the fusion
barrier but fail to evolve inside the fission saddle point and reseparate. Thus 
the process
differs from fission fragments emitted from a fully equilibrated compound
nucleus. To distinguish these two processes in a low statistics RIB
experiment is very difficult. Nevertheless,
one-dimensional barrier penetration models and coupled-channel calculations
give results
for capture inside the fusion barrier which includes quasifission for
heavy systems. Here, the fusion cross section actually
refers to the capture cross section.

\begin{figure}[th]
\centerline{\psfig{file=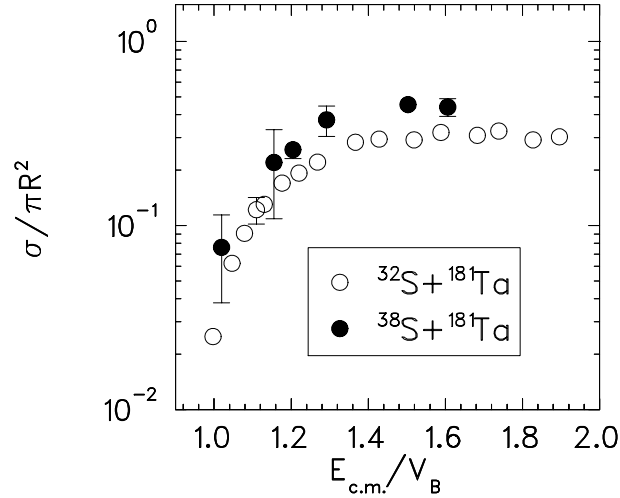,width=8cm}}
\caption{Reduced fusion-fission excitation functions for $^{38}$S+$^{181}$Ta
(closed circles) and $^{32}$S+$^{181}$Ta (open circles). \label{fg:s38ta}}
\end{figure}

The excitation functions for $^{32}$S and $^{38}$S induced fusion are compared
in Fig.~\ref{fg:s38ta} by reduced cross section and reduced energy as
described in the previous section. The
fusion cross sections for $^{38}$S+$^{181}$Ta are generally larger than those
for $^{32}$S+$^{181}$Ta. It is noted that the comparison is presented
differently in Ref.~\cite{zy97,zy01} where the reduced cross section and
reduced energy are calculated
with the experimentally extracted barrier radius and barrier
height, respectively. In that case, the two excitation functions coincide.
The barrier height $V_{b}$ and barrier radius R$_{b}$ extracted by fitting
the high energy cross sections to the classical limit
$\sigma=\pi R^{2}(1-V_{b}/E)$ show that the barrier height was reduced
significantly, by 5.9
MeV, and the barrier position shifted by 1.8 fm for $^{38}$S with respect to
$^{32}$S.\cite{zy01}
According to the Bass model, the barrier height decreases by
3.4 MeV and the barrier radius increases by 0.4 fm. The larger
barrier shift may be attributed to the deformation of $^{38}$S 
($\beta_{2}$=0.246)\cite{sc96}. This deformation, not accounted for in the
Bass model, may result in the difference shown between Fig.~\ref{fg:s38ta}
(enhancement) in this paper and Fig.~4 in Ref.\cite{zy97} (no enhancement).

The Q-value for $^{38}$S induced fusion is 6.3 MeV lower than that for
$^{32}$S induced fusion. The excitation energy of the compound nucleus is,
therefore, 12.2 MeV lower at the barrier. As the excitation energy of the
compound nucleus decreases, the total number of neutrons evaporated decreases,
however, the strength of lower multiplicity neutron evaporation channels
actually increases. This is a very valuable piece of information.
A colder compound nucleus formed in neutron-rich radioactive nucleus induced
fusion will increase the production yield of neutron-rich heavy elements.

\subsection{$^{132}$Sn beam}
The advent of accelerated $^{132}$Sn for experiments is important for
nuclear physics since $^{132}$Sn is a doubly magic nucleus
(Z=50, N=82). It has a
large N/Z ratio (1.64) and has eight extra neutrons compared to the heaviest
stable Sn isotope, $^{124}$Sn. The fusion-evaporation excitation functions
of $^{64}$Ni on even Sn isotopes from A=112 to 124 have been
measured.\cite{fr83} Being
able to extend the Ni+Sn systems to the very neutron-rich $^{132}$Sn+$^{64}$Ni
system is very exciting.

The $^{132}$Sn beams were
produced with the ISOL technique at the HRIBF. The short-lived $^{132}$Sn
($t_{1/2} = 39.7$ s) is a product of
proton induced $^{238}$U fission. Since most of the mass 132 isobars
produced is $^{132}$Te, sulfur was introduced into the ion source to purify
the beam.\cite{st03} This relies on the fact that sulfur and tin form
chemical compounds at a much higher rate than other produced mass 132 isobars.
The SnS$^{+}$ molecular ions were subsequently
broken up in a charge exchange cell. In this way, the composition of the beam
consisted of 96\% $^{132}$Sn. The small amount of $^{132}$Te had negligible
effects on the measurement because the higher Z gives a higher Coulomb barrier
and fusion is, therefore, suppressed. The Sn$^{-}$ ions were accelerated
in the 25 MV tandem electrostatic accelerator. The average beam intensity for
the experiment was 2$\times 10^{4}$ pps with a maximum of 5$\times 10^{4}$ pps.
Since at energies near and below the Coulomb barrier the predominant decay mode
of the compound nucleus is particle evaporation, residue cross
sections were measured. This is an inverse kinematics reaction, therefore,
the residues
are forward focused. A combination of time-of-flight detectors and an
ionization chamber placed at zero degrees provided high efficiency for residue
detection. Cross sections of the order of mb can be measured.

\begin{figure}[th]
\centerline{\psfig{file=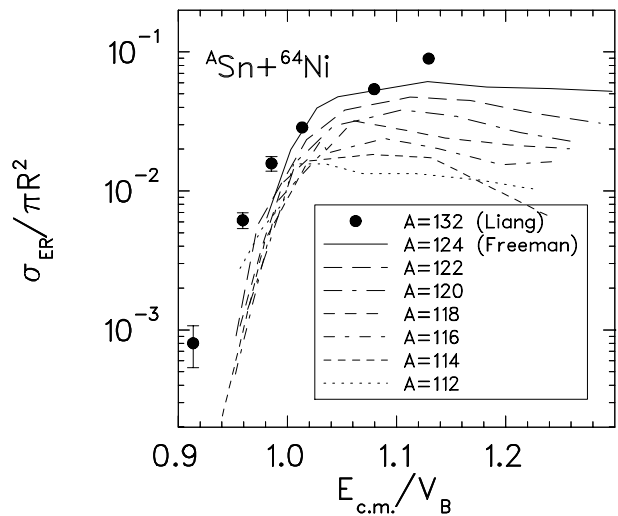,width=8cm}}
\caption{Reduced evaporation residue excitation function for
$^{64}$Ni+$^{A}$Sn. The filled circles are for reactions induced by $^{132}$Sn.
\label{fg:sn132ni}}
\end{figure}

\begin{figure}[th]
\centerline{\psfig{file=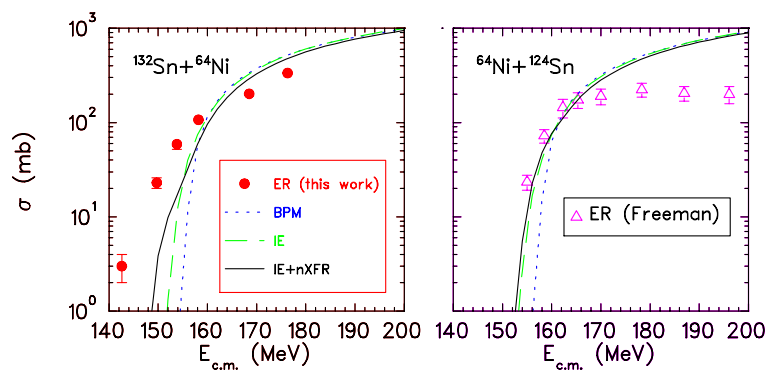,width=10cm}}
\caption{Comparison of measured ER excitation functions with
coupled-channel calculations. The left panel is for $^{132}$Sn+$^{64}$Ni
and the right panel is for $^{64}$Ni+$^{124}$Sn\protect\cite{fr83}.
The measured ER cross
sections are shown by the filled circles and open triangles for
$^{132}$Sn+$^{64}$Ni and $^{64}$Ni+$^{124}$Sn, respectively.
One-dimensional barrier penetration model prediction, coupled-channel
calculations including
inelastic excitation, and inelastic excitation and transfer are shown by
the dotted, dashed and solid curves, respectively.
\label{fg:sn132cc}}
\end{figure}

The evaporation residue excitation function for $^{132}$Sn+$^{64}$Ni is shown
in Fig.~\ref{fg:sn132ni} in reduced coordinates.\cite{li03b}
The cross sections for stable
even Sn isotopes are shown in the same figure for comparison. As can be
seen, the $^{132}$Sn induced fusion is very much enhanced at sub-barrier
energies. Coupled-channel calculations were performed to compare with the
measured data. Since statistical model calculations predicted that fission is
negligible at E$_{cm} \leq$ 160 MeV, the evaporation residue cross sections
are taken as fusion cross sections. Coupled-channel calculations including
projectile and target excitation to their first 2$^{+}$ and 3$^{-}$ states
are in fair agreement with $^{64}$Ni+$^{124}$Sn data but fail to reproduce
$^{64}$Ni+$^{132}$Sn data, as shown in Fig. \ref{fg:sn132cc}. It is noted that
$^{64}$Ni+$^{124}$Sn has only
one transfer channel, ($^{64}$Ni,$^{66}$Ni), with a positive Q-value. For
$^{64}$Ni+$^{132}$Sn, the Q-values for $^{64}$Ni picking up two to six
neutrons are positive. Coupled-channel calculations including inelastic
excitation and transfer reactions reproduces well the $^{64}$Ni+$^{124}$Sn
data.
But there is still large discrepancy between the calculation, shown by the
solid curve in Fig. \ref{fg:sn132cc}, and measurement
for $^{64}$Ni+$^{132}$Sn. It should be pointed out that the form factors
implemented for the multinucleon transfer\cite{li04} were extrapolated from the
$^{58}$Ni+$^{124}$Sn measurements.\cite{ji98} Further
development of the calculation and measurement of nucleon transfer are
necessary for better understanding the enhancement observed here.
The $^{64}$Ni+$^{124,132}$Sn comparison is very similar to that of
$^{40}$Ca+$^{90,96}$Zr.\cite{ti98} There is no transfer channel with positive
Q-value
in $^{40}$Ca+$^{90}$Zr, but several neutron transfer channels with positive
Q value are in $^{40}$Ca+$^{96}$Zr. Coupled-channel calculations considering
inelastic excitation reproduce well the $^{40}$Ca+$^{90}$Zr data. When
inelastic excitation and transfer were coupled, the calculations
underpredicted the $^{40}$Ca+$^{96}$Zr data. However, only
simultaneous nucleon transfer was considered in their calculations
and sequential nucleon transfer
may be more important.\cite{ti98} Moreover, Stelson's neutron flow
mechanism may play a role in $^{40}$Ca+$^{96}$Zr because the barrier
distribution
is broader and flatter as compared to that of $^{40}$Ca+$^{90}$Zr. It is
conceivable that the neutron flow mechanism can become more dominant than
the coupling to surface modes in very neutron-rich nucleus induced fusion.
When the $^{132}$Sn beam intensity is greater in the future, it will be
very useful to perform
high precision excitation function measurements to study the barrier
distributions.

In the $^{38}$S and $^{29,31}$Al experiments, no further fusion enhancement
relative to the stable isotope induced reaction
was found, but large enhancement was observed in
the $^{132}$Sn experiment. It is noted that
in the former two experiments the neutron-rich nucleus is the light reactant
whereas in the latter experiment, the neutron-rich nucleus is the heavy
reactant. Moreover, $^{132}$Sn has eight extra neutrons compared to the
heaviest stable Sn isotope whereas $^{38}$S and $^{31}$Al have two and four
extra neutrons compared to the heaviest stable S and Al isotopes, respectively.

In fusion induced by neutron-rich nuclei such as $^{38}$S, the
lowering of the excitation energy can increase the cross section for lower 
multiplicity neutron evaporation channels. If the fusion is further enhanced as
seen in the $^{132}$Sn induced fusion, the cross section for lower multiplicity
neutron evaporation channels can be further
increased. There could be an advantage to using neutron-rich RIBs to produce
heavy elements if the intensity is sufficiently high.
Experiments using more neutron-rich radioactive nuclei are underway to look
for systematic trends.
These new experiments will provide more clues for
understanding the mechanisms for fusion enhancement. As the nuclei become
more neutron-rich, the neutron binding energy decreases and neutron breakup
will start to play a role. Whether neutron breakup in heavy-ion reactions
influences fusion the same way as light-ion reactions and the implication on
heavy element production will be interesting to study.
 
\section{Concluding Remarks}
\subsection{Present situation}
From the analysis of the various results presented in this review it is clear
that presently they suffer from the lack of good statistics as compared with
those obtained with stable beams which are usually at least three orders of
magnitude more
intense. The situation will remain like this for sometime until new
dedicated RIB facilities will be in operation.\cite{ri00,eu03,riken} These
first generation experimental data give still valuable results, but one should
only expect to get hints from them rather than clear indications as with
stable beams.
For the moment, the main indications, keeping in mind the $^{6}$He results,
are that for light nuclei with loosely bound neutrons, there is not such a
large sub-barrier fusion enhancement as predicted by many theories. However,
there is a new effect which appears in a strong channel(s) where the light
projectile fragment(s) are trapped into the target by the nuclear field in a
nonconventional transfer process that could be called
transfer/stripping-breakup.
Such a process seems to be weaker with light projectiles when
they have loosely bound protons as in $^{17}$F. This different behavior
could be tentatively explained by invoking different effects of the
polarization potentials for the two projectiles. The details of such a process,
which seem to be strongest at the barrier (see Fig.~\ref{fg:he6comm}), have to
be further investigated theoretically and experimentally.

With medium mass RIBs, no fusion enhancement is expected from the halo
structure since the beams that are available today are not near the drip-lines.
The relevant point could be the
large neutron excess of some specific RIBs like $^{132}$Sn.
The data are, for the moment, quite scarce and scattered throughout the
nuclear chart so it is nearly impossible to see any trend. Only for the
heaviest
system discussed in this paper is sub-barrier fusion enhanced. This may
be attributed to the large neutron excess of the projectile, $^{132}$Sn, which
has eight neutrons more than the heaviest stable $^{124}$Sn. It
is conceivable that neutron transfer plays an important role in enhancing
fusion yields. But the simplified treatment of transfer channels in
the theory fails to predict such an enhancement. More
systematic data focused on specific systems will be necessary for heavy
systems in order to identify a trend in fusion.

\subsection{Perspectives}
Nuclear reactions are often used as tools for probing nuclear structure.
As more species of RIBs become available and further away from
stability, fusion induced by neutron-rich radioactive nuclei could be used to
explore the properties of neutron skin and neutron halo in heavier nuclei.

In very heavy reaction systems, the extra-push energy is required for complete
fusion.\cite{sw81} The extra-push energy depends on the effective fissility
parameter,
a measure of the Coulomb repulsion against the 
nuclear surface tension, of the system. How
the extra-push energy affects fusion induced by very neutron-rich radioactive
nuclei is an important research subject. It can provide information on
whether it is practical to use such beams for producing
heavy elements. Another quantity related to heavy element production is
the survival probability, which correlates with the fission barrier.
The systematics of fission barrier heights is obtained from stable
and proton-rich nuclei\cite{si86}. With shell corrections, fission barriers
for neutron-rich heavy elements were predicted. The location of shell
closure in very heavy nuclei may be probed by fusion studies using
neutron-rich radioactive beams. To better
understand the reaction dynamics, measuring evaporated particles from
neutron-rich radioactive nucleus induced fusion will be useful. One can also
learn about the level density of a neutron-rich compound nucleus.

It has been demonstrated that barrier distributions can be extracted from
high precision fusion excitation function measurements.
The distribution of barriers 
can reveal the signature of channel couplings. It identifies the important
channels which contribute to the sub-barrier fusion enhancement. With the RIBs
available today, it is not practical to perform such measurements because the
intensity is orders of magnitude too low. Some proposed new RIB facilities
are designed to deliver beams of intensity comparable to stable beams. These
new facilities will allow the barrier distribution to be measured in a
reasonable period of beam time.

While we are devoting a great deal of effort to RIB experiments, we should not
neglect stable beam experiments. We have seen that the important
results obtained with $^{6,7}$Li and $^{9}$Be shed light on the
influence of breakup on fusion. As we move further away from
stability, the beam intensity is expected to be much lower. Experimental
apparatus will be more complicated and experiments will take longer
times. Stable beams
will have to be used to setup and calibrate equipment. Stable beam experiments
may also provide valuable information to help us understand the results of
RIB experiments which frequently suffer from poor statistics.

In the next few years, many facilities will provide more varieties of RIBs
with energies above the Coulomb barrier. The projected intensity for some of
the beams will reach above 10$^{8}$ pps. More fusion experiments will be
performed and we expect to see some measurements with very good statistics.
Reaction channels that are important to sub-barrier fusion enhancement, such
as nucleon transfer and inelastic excitation can
be measured with higher intensity beams and dedicated apparatus. At the same
time, experiments will push towards using beams further away from the
stability. Progresses in theoretical treatment of fusion involving RIBs will
be made with new experimental results. Furthermore, there are several new and
powerful RIB facilities\cite{ri00,eu03} that are
under consideration. When they become a reality, more exciting fusion
experiments and results can be expected.
%\begin{figure}[th]
%\centerline{\psfig{file=ijmpef1.eps,width=5cm}}
%\vspace*{8pt}
%\caption{A schematic illustration of dissociative recombination. The
%direct mechanism, 4m$^2_\pi$ is initiated when the molecular ion
%S$_{\rm L}$ captures an electron with kinetic energy.}
%\end{figure}

%Previously published material must be accompanied by written 
%permission from the author and publisher.

%\begin{table}[pt]
%\tbl{Comparison of acoustic for frequencies for piston-cylinder problem.}
%\begin{tabular}{@{}cccc@{}} \toprule
%Piston mass & Analytical frequency & TRIA6-$S_1$ model &
%\% Error \\
%& (Rad/s) & (Rad/s) \\ \colrule
%1.0\hphantom{00} & \hphantom{0}281.0 & \hphantom{0}280.81 & 0.07 \\
%0.1\hphantom{00} & \hphantom{0}876.0 & \hphantom{0}875.74 & 0.03 \\
%0.01\hphantom{0} & 2441.0 & 2441.0\hphantom{0} & 0.0\hphantom{0} \\
%0.001 & 4130.0 & 4129.3\hphantom{0} & 0.16\\ \botrule
%\end{tabular}}
%\begin{tabnote}
%Table notes
%\end{tabnote}
%\begin{tabfootnote}
%\tabmark{a} Table footnote A\\
%\tabmark{b} Table footnote B
%\end{tabfootnote}
%\end{table}

\section*{Acknowledgements}
The authors wish to thank A. di Pietro, M. Trotta, N. Alamanos, N. Keeley,
J. J. Kolata, P. E. Mueller, R. Raabe, and D. Shapira for kindly providing
some of their data relevant for
the preparation of this review and/or for critical reading of this manuscript.
Research at the Oak Ridge National Laboratory is
supported by the U.S. Department of Energy under contract DE-AC05-00OR22725
with UT-Battelle, LLC.

\end{document}